\documentclass{xsurv_article}

\input{psfig.sty}

\def\etal{{\it et al.~}}
\def\cm2-sec{\ cm$^{2}$-sec}
\def\la{\hbox{\rlap{$<$}\lower.5ex\hbox{$\sim$}\ }}
\def\ga{\hbox{\rlap{$>$}\lower.5ex\hbox{$\sim$}\ }}
\def\deg{$^{\circ}$~}
\def\arcmin{$\,^\prime$~}

\begin{document}  
 
\begin{titlepage}
\setcounter{page}{1}
\makeheadline

\vspace*{-1cm}
 
\title{Surveying the Hard X-ray Sky: Imaging in Space and Time}  
 

\author{{\sc Jonathan E. Grindlay}, Cambridge, USA \\
\medskip
{\small Harvard-Smithsonian Center for Astrophysics} 
} 

\date{Received; accepted } 
\maketitle

\summary
 One of the few remaining astronomical bands (factor of $\sim$10 in 
energy range) still without an all-sky imaging survey is the hard 
x-ray band (10-600 keV). This is in spite of sensitive imaging all-
sky surveys already conducted at soft x-ray  (0.2-2 keV; ROSAT) and 
soft/hard $\gamma$-ray ($\sim$750 keV - 10 GeV; COMPTEL/EGRET)  
energies and imminent 
for medium x-ray energies (2-10 keV; ABRIXAS). 
A hard x-ray imaging survey conducted with wide-field coded aperture 
telescopes allows both high sensitivity (and spatial/spectral resolution) 
and broad temporal coverage. We derive a generalized survey sensitivity/temporal 
parameter, Q, and compare previous and planned hard x-ray surveys 
with the proposed EXIST mission. Key scientific objectives that 
could be addressed with the enhanced capability of EXIST are outlined. 
END

\keyw
x-ray surveys; x-ray binaries; black holes; neutron stars; AGN
END

\AAAcla
END
\end{titlepage}

\kap{Introduction}
All sky surveys in the soft-medium x-ray band, defined here as 0.2 - 10 keV, 
are  most effectively carried out with focussing x-ray telescopes executing 
scanning missions. The premier examples are of course ROSAT (0.2-2 keV; 
Trumper 1983) and the planned (1999 launch) ABRIXAS mission (2-10 keV; 
Staubert, these proceedings). At hard x-ray energies, defined here as being 
10-600 keV to overlap with the focussing soft-medium surveys and to both 
study the non-thermal universe and to extend above the 
natural astrophysical break point of 511 keV 
positron emission, the techniques and capabilities 
of an all sky survey change dramatically. Instead of being  
a narrow field (instantaneous) 
survey limited in temporal coverage, 
a hard x-ray survey can be executed as a 
wide field scanning survey sensitive to a broad range of timescales.
The wide field ($\sim$30\deg) but pointed soft/hard $\gamma$-ray surveys 
carried out with COMPTEL/EGRET ($\sim$750 keV-10 GeV) on CGRO  
are limited by the pointing, rather than survey, duration.

We consider a simple general parameterization for x-ray surveys (soft-hard) 
which can (and should) survey both spatial and temporal images, ideally all sky. 
We compare the parameters of the principal missions  already carried out or 
now in final planning stages and extend this to the proposed EXIST 
mission concept (Grindlay et al 1995, 1997). We then summarize briefly 
the key scientific objectives of EXIST and then outline the wide-field 
coded aperture imaging telescope and detector system which would allow 
such a mission to be conducted.

\kap{Generalized Surveys}

Surveys are  essential for moving beyond the study of a few objects pre-selected 
for their likely emission or properties based on other bands, and they 
allow the discovery of entirely new classes of objects and astrophysical 
phenomena. Population and evolution studies require the complete samples of surveys. 

Generally surveys are either spatial (e.g. deep images, mosaiced 
to cover a particular field or the whole sky) or are temporal (e.g. monitoring 
of time variability of a source or collection of sources). For 
maximum sensitivity, 
in either flux or temporal resolution, imaging is essential. Source confusion 
is the familiar limit for a given beam size, or imaging 
resolution element, r, but 
is usually not the limiting factor for non-imaging (e.g. collimated, or 
scanning) x-ray or gamma-ray surveys. Rather, these are 
instead usually limited in flux 
by either short exposure, high backgrounds, or both. Thus a spatial survey 
of total coverage or solid angle (fov), F 
(vs. instantaneous telescope fov, G), has M = F/r$^2$ pixels for which a sensitivity 
limit of minimum detectable source flux, f, is achieved per pixel and per survey 
total exposure time. Similar considerations  
apply to temporal surveys, where the image 
resolution element (time resolution, t) 
and image fov (time duration, T), define 
the number, N = T/t, of temporal pixels 
in the temporal image that can be formed with  flux sensitivity, g,  per pixel. 
For each spatial image pixel, the temporal flux 
sensitivity limit, g, is usually significantly 
greater than the spatial flux limit, f, since 

\begin{equation}
g  \sim f \cdot \sqrt{\rm{N}}
\end{equation}

\noindent
for all but periodic source behaviour. In the case of a survey for known periodic 
sources (e.g. monitoring of x-ray pulsars), the above sensitivity estimate 
would apply with N replaced by p, the number of phase bins containing the pulsed 
signal. 
   
An ideal survey would maximize M and N, which are the effective survey size/sensitivity 
parameters m = M/f and n = N/g. Surveys can then be compared by their 
tracks in a normalized m,n plane. We may also define an overall survey quality 
factor, Q, which combines the spatial and temporal parameters by their harmonic 
mean:

\begin{equation}
Q = \sqrt{\rm{m} \cdot \rm{n}} = \rm{M}^{1/2} \cdot \rm{N}^{1/4}/f 
\end{equation}

\noindent
where we have used Equation 1  for the relative 
sensitivities f and g appropriate 
to the more general case of a survey sensitive to bursts and transients of 
unknown timescale. 

\kap{Previous Hard X-ray Surveys} 
It is remarkable that the hard x-ray band (10-600 keV) has 
been  neglected for the two decades since the only survey conducted by 
the HEAO-A4 experiment. This pioneering first survey was  limited 
in  sensitivity (reaching  $\sim$50 mCrab at 13-180 keV) and 
both angular resolution ($\sim$3\deg; set by the non-imaging detector 
with 1.5\deg $\times$ 20\deg collimator or effective G $\sim$6\deg)  
and temporal coverage and resolution (each source 
observed, typically, for only $\sim$1 week each 6 months, for three such 
observations).  Some  70 sources were detected, all known from 
previous soft-medium x-ray surveys (e.g. UHURU and HEAO-A1),  and 
of these only 14 were detected in the 80-180 keV band (Levine et al 1984). 
These objects were (understandably) the brightest in their respective 
classes, including Crab, Cyg X-1, Cen A, NGC 4151 and 3C273, and no new 
classes of object were discovered. 

Only one other all-sky survey at hard x-ray energies (20-100 keV) 
has now  effectively been 
carried out (although not yet  processed) using the BATSE experiment:  
A galactic plane survey (b = $\pm$20\deg) is in progress 
and preliminary results have been reported (Grindlay et al 1996) 
using the automated occulation-image scanning analysis system 
CBIS (CfA BATSE Image Scan) developed at CfA. Extension 
to a full sky survey could be done for similar total sensitivity ($\sim$50 mCrab), 
angular resolution ($\sim$1\deg) and temporal resolution ($\sim$1 orbit = 1.5h). 
\vspace*{0.3cm}

\pagestyle{empty}

\setcounter{secnumdepth}{5}

\centerline{Table 1: Chronological Summary of Hard X-ray Surveys}
\vskip 3pt
\small
\begin{center}
\begin{tabular}{lrlllllll}
Mission &  $\Delta$E (keV) & Det./Tel. Sys.& F, G (\deg) & r (\deg)& T (d)& 
t (h) & f (mCrab) & Q   \\ \hline
&&&&& \\

HEAO-A4  & 13-180  & NaI; scan collim. & AS, 6   & 3 & 1e+3 & 0.5 & 50 & 20 \\
BATSE    & 20-100  & NaI; occultation  & AS, AS/2 & 1 & 2e+3 & 1.5 & 50 & 54 \\
ABRIXAS  & 2-10    & CCD; focussing    & AS, 0.7  & 0.03 & 1e+3 & 1.5 &0.2 &3.7e+5 \\
INTEGRAL & 15-200 & CdTe; coded mask  & GP, 20 & 0.2 & 1.e+3 & 2e-4 & 3 & 1.5e+4 \\
EXIST    & 10-600 & CdZnTe; coded mask & AS, 40 & 0.2 & 270 & 2e-4 & 0.3 & 2.5e+5 \\

\end{tabular}
\end{center}
\small
Notes: AS = all sky: F = 40,000 deg$^2$; GP = gal. plane: 
F $\sim$ 7200 deg$^2$. See text for definitions of G, r, T, t, f and Q.
\vskip 3pt

\normalsize
Otherwise there are at present only two planned surveys which 
partially overlap with a full all-sky hard x-ray imaging 
survey: the ABRIXAS survey (2-10 keV; Staubert, these proceedings) 
and the galactic plane survey to be conducted with INTEGRAL 
(15 keV-10 MeV; Ubertini et al 1996). These missions are planned 
for launches in 1999 and 2001, respectively. 
In Table 1 we summarize these various missions by their key 
parameters as well as the survey quality parameter Q  defined 
above (Eq. 2). EXIST complements ABRIXAS and greatly extends  
the energy range and temporal coverage at comparable total sensitivity. 
 
\kap{EXIST Concept and Capability}
  
The Energetic X-ray Imaging Survey Telescope (EXIST) was 
selected (April 1995) as one of  the 27  
New Mission Concept (NMC) studies  for  a future NASA  
astrophysics mission. It would conduct the first  
hard x-ray imaging all sky survey with a 
sensitivity some 100$\times$ greater than the 
HEAO-A4 experiment in 1977-79   
(Levine et al 1984). The need, and priority, for such
an all sky imaging hard x-ray survey mission has been pointed out 
in the recent report of the NASA Gamma-Ray Program Working Group. 
An overall description of the initial and proposed EXIST   
concepts is given by Grindlay et al (1995, 1997) (and see 
{\it http://hea-www.harvard.edu/EXIST/EXIST.html}).   

EXIST would incorporate two coded aperture telescopes, each 
with fov = 40\deg $\times$ 40\deg (above $\sim$40 keV; at 10-30 keV 
a low energy collimator would restrict the FOV to 3.5\deg in one 
dimension), with a  combined fov =  40\deg $\times$ 80\deg yielding 
60\% sky coverage each orbit and full sky each 50d orbital precession period.
A Cd-Zn-Te (CZT) pixellated detector array of total area of 2500 cm$^2$ is 
at the focal plane of each telescope.   

\begin{figure*}[t]  
\centering  
 
\hspace*{-3.2in}  
\psfig{figure=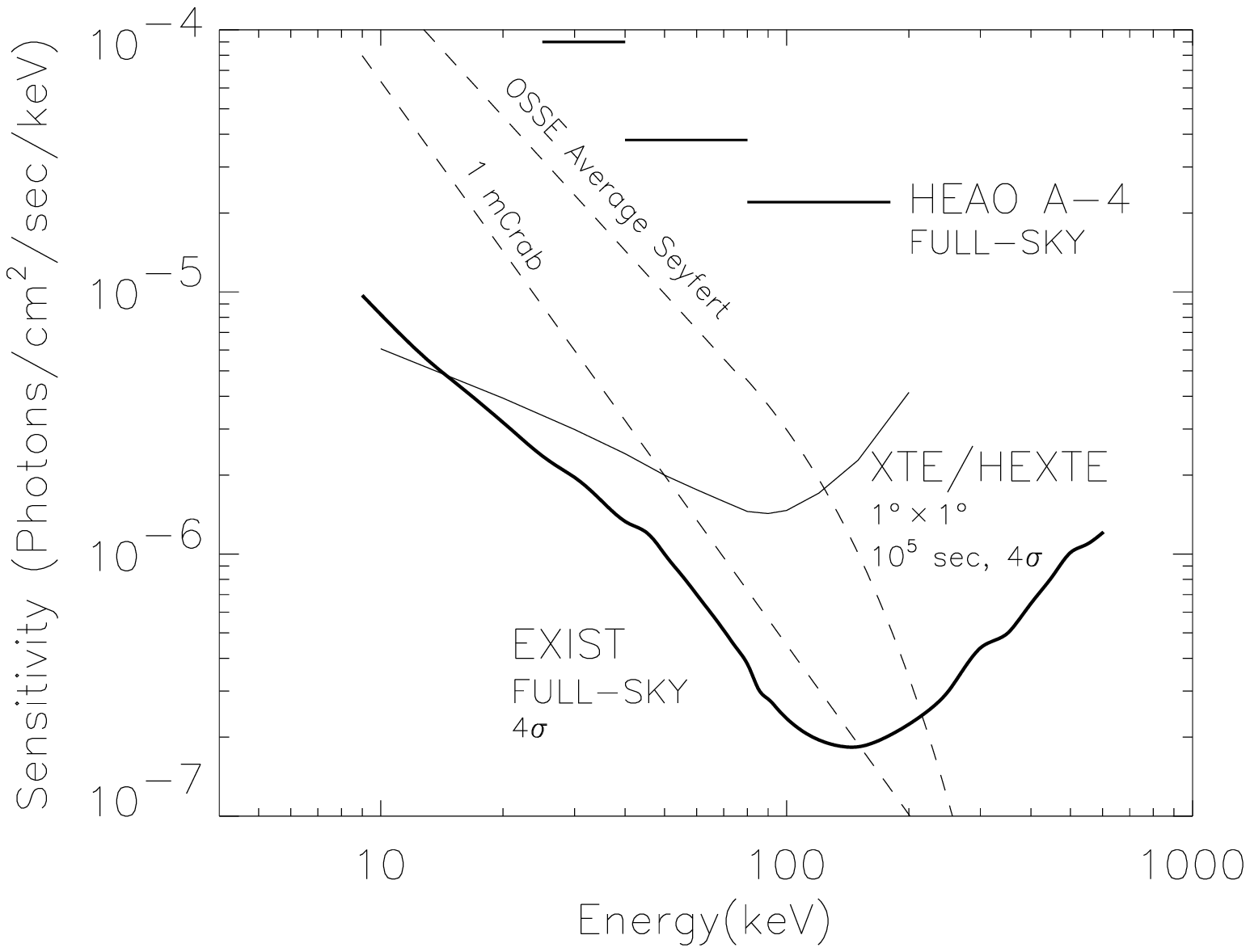,height= 3.in,width=3.7in,angle=0}  
  
\vspace*{-3.in}  
  
\hspace*{3.35in}  
\psfig{figure=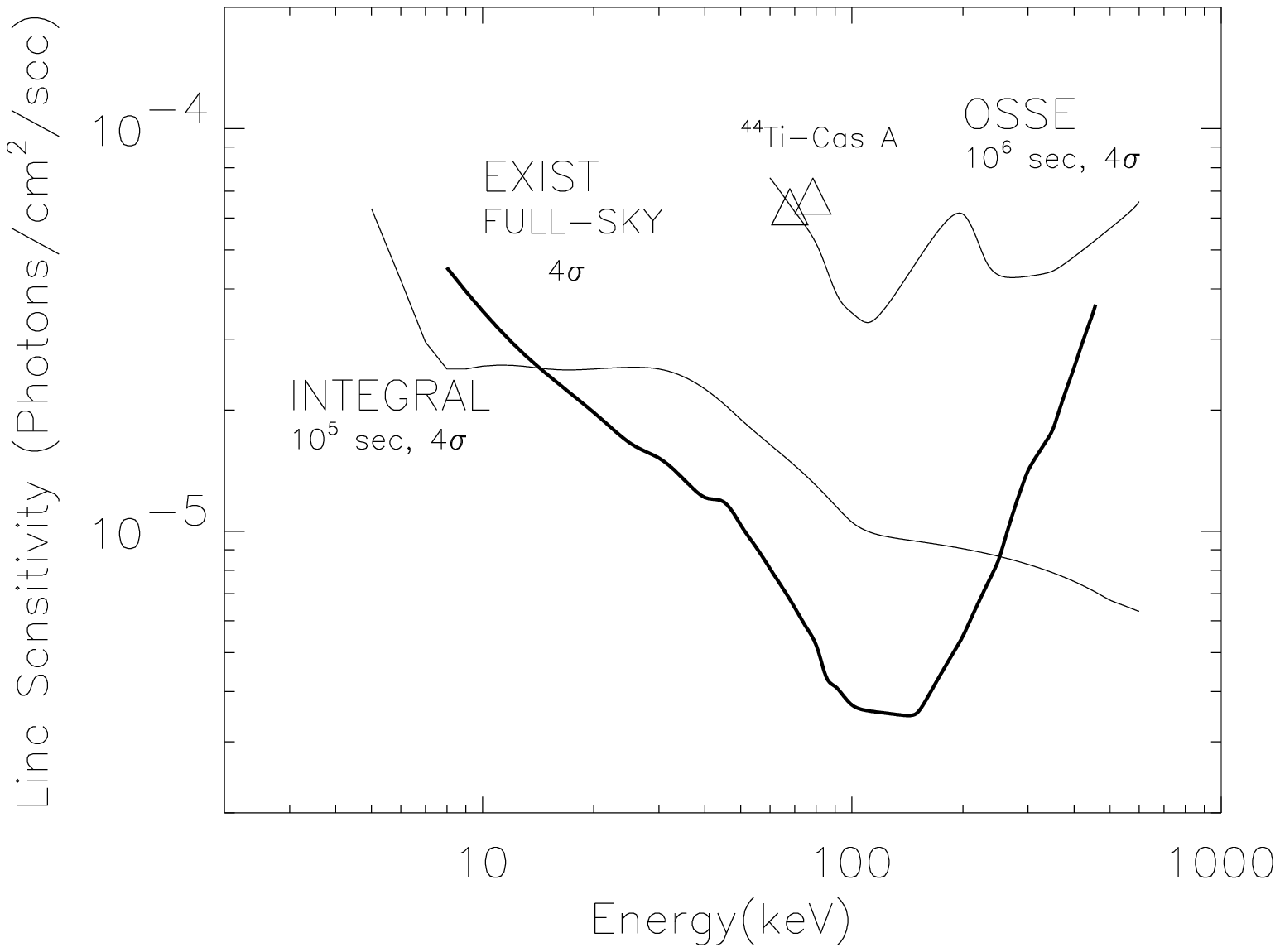,height= 3.in,width=3.7in,angle=0}\\  
  
\caption{Sensitivity of EXIST (MIDEX) for continuum (left)  
and narrow line (right) spectra vs. other missions. }

\end{figure*}  
  
Because of its very large fov and large area detectors with   
high intrinsic resolution (both spatial and spectral), EXIST 
could reach the unprecedented all-sky sensitivity  
shown in  Figure 1.   
The sensitivity plots are for EXIST for its proposed   
9-month all-sky survey  which would allow total integration times of   
$\sim$10$^6$ sec for any source. The mission 
could then be operated as a pointed (observatory) 
for additional exposure and higher sensitivity on selected high priority 
targets (e.g. M31 GRB/SGR 
 and BHC surveys; galactic 
bulge survey). EXIST could be proposed for MIDEX or 
as a Space Station attached payload. 
 
\kap{Scientific Objectives: All Sky Survey and Monitoring} 
 
\sect{Surveys}

{\it Hard x-ray spectra and luminosity function of AGNs:} Active galactic  
nuclei (AGN) are now measured by OSSE to have hard spectra with breaks  
typically in the 50-100 keV range for Seyferts (cf. Zdjiarski et al 1995)  
and with multi-component or non-thermal spectra extending to  
higher energies likely for the  
blazars. EXIST will have an all-sky sensitivity 
some 10$\times$ better than  
that needed to detect the ``typical'' Seyferts seen with OSSE.  
More than 1000 AGN should be detected in the all sky survey and   
EXIST has the required sensitivity in   
the poorly explored 30-200 keV band to measure 
accurate spectra for all known AGN detected   
with Ginga or with the Einstein slew survey.  
 
A major objective for AGN studies and surveys is the 
detection and inventory of  heavily obscured or self-absorbed AGN.  
Such objects, primarily Seyfert 2's but also 
including (some) star-formation  
galaxies, are now being discovered in pointed observations with SAX  
and will also likely 
be discovered with the focussing  
ABRIXAS all sky survey. 
However, the most heavily obscured objects, with absorbing  
column densities \ga10$^{24-25}$cm$^{-2}$ yielding low energy cutoffs  
in the 5-10 keV range, will be more readily detected with EXIST. This   
will yield the first measure of the luminosity function of  
obsured AGN, which are likely significant for the x-ray background.  
 
{\it Survey for black hole and neutron star compact binaries.}  
Studies of compact objects over a wide range of   
timescales and luminosity are possible 
throughout the Galaxy. A deep galactic survey  
for x-ray binaries containing  black holes vs. neutron 
stars and pulsars will allow the  
relative populations of black holes in the Galaxy to be constrained.   
All previous galactic hard x-ray surveys have been constrained to 
the brightest decade in source flux (and luminosity); EXIST would  
extend this 1-2 decades deeper, allowing 
spectral studies. Whereas the INTEGRAL galactic 
plane survey(s) will also make great strides, the EXIST (all sky) survey 
would be more sensitive and not be limited to the central $\sim\pm$10\deg 
of galactic latitude covered by the smaller fov of INTEGRAL. 
 
{\it Emission line surveys: hidden supernovae via $^{44}$Ti  
emission and 511 keV sources:}  
The array of CZT imaging detectors proposed for EXIST  
achieves high spectral resolution (e.g. \la5\% at 60 keV)   
enabling emission line surveys.   
The decay of $^{44}$Ti (lines at 68 and 78 keV) with long (68 y)  
halflife allows a search for the long-sought population of  
obscured supernovae in the galactic plane at sensitivities   
significantly better than the possible detection of Cas-A (cf. Figure  
1). Similarly, 511 keV emission from black hole   
binaries (or AGN) can be searched for (e.g. in transient outbursts) 
and imaged with better sensitivity than OSSE (cf. Figure 1).   
  
{\it Study of the diffuse hard x-ray background:} 
The spectra of a significant sample of AGN will test the AGN   
origin of the diffuse background  for the poorly  
explored hard x-ray band. Because the background measured by  
the EXIST detectors below 100 keV is dominated by the cosmic diffuse  
spectrum, its isotropy and fluctuation spectrum can be studied   
with much higher sensitivity than before.

\sect{Monitoring} 

EXIST would survey 60\% of the sky each orbit anc accumulate $\sim$10 orbits/day 
(allowing for SAA, etc.) 
$\times$ \ga10 min exposure/orbit or \ga6000 sec/day for each source 
observed. This yields a daily flux sensitivity (30-100 keV) 
of $\sim$1-2 mCrab, sufficient 
for the brightest AGN and essentially all known accretion-powered 
binaries in the Galaxy. Pulsar timing allows even fainter flux limits, 
as demonstrated with the extensive BATSE 
monitoring project.  Over one sky survey epoch ($\sim$50d), 
each source is observed 
for \ga25d, giving $\sim$0.3mCrab limits for $\sim$month timescales.  

{\it Faint hard x-ray transients: black hole population in Galaxy:} 
The sensitivity to $\sim$1-10d transients is \ga30 $\times$ better than 
BATSE, so that the low resolution occultation-imaging 
survey for faint transients being conducted with BATSE (cf. 
Grindlay et al 1996) can be extended to correspondingly lower 
outburst luminosities or greater source distances. With a 
1-10d sensitivity of $\sim$1 mCrab, BH transients  can be 
detected with their characteristic peak luminosities of 
$\sim10^{37-38}$erg/s (10-100 keV) out to 100 kpc, enabling 
the first BH-transient survey of the LMC/SMC. 

{\it Monitoring and Study of X-ray Pulsars:}  
The measurement and monitoring of spin periods, pulse 
shapes and luminosity/spectra   
of a large sample of accretion-powered pulsars would 
extend the BATSE sample  of Bildsten et al (1997)  
to the entire sample (\ga30) of known   
accretion-powered pulsars. The high spectral resolution 
of the CZT detectors on EXIST would allow high sensitivity 
studies of cyclotron features in pulsar spectra, greatly 
extending current RXTE/HEXTE studies of relatively few 
objects to a much larger sample. 
 
{\it Studies of  Gamma-ray Bursts:}   
EXIST would have a GRB sensitivity approximately 20$\times$ that of  
BATSE so should detect GRBs overall  
at about 1/2 the rate, or $\sim$0.5/day, as BATSE with its   
much larger fov but reduced sensitivity. GRBs would be  
located to \la1-5\arcmin positions, thereby providing definitive tests  
of repeaters. Burst positions and spectra would be brought down  
in real time for rapid followup studies of GRB afterglows and identifications.  
 
\acknowledgements
The EXIST Concept Study is  
supported in part by NASA grant NAG8-1212.

\refer
\rf{Bildsten, L. \etal 1997, ApJ Suppl., in press} 
\rf{Grindlay, J., Prince, T., Gehrels, N.., Tueller, J., Hailey, C.   
\etal 1995,  Proc. SPIE,  2518, 202}
\rf{Grindlay, J., Prince, T., Gehrels, N.., \etal 1997, in Proc. ASM
Wkshp, Tokyo, in press}
\rf{Grindlay, J. Barret, D., Bloser, P. \etal 1996, A\&A, 120, 145}
\rf{Levine, A.M. \etal 1984,  ApJ Suppl., 54, 581} 
\rf{Trumper, J. 1983, Adv. Sp. Res., 2(4), 241}
\rf{Ubertini, P. \etal 1996, Proc. SPIE, 2806, 246}
\rf{Zdziarski, A., Johnson, W., Done, C. \etal 1995, ApJ, 438, 63}
 

\vspace*{-0.5cm}
%
\addresses
\rf{Jonathan E. Grindlay, Harvard Observatory, 
Harvard-Smithsonian Center for Astrophysics,\\ 
60 Garden St., Cambridge, MA 02138, USA, e-mail: 
jgrindlay@cfa.harvard.edu}
END
%

\end{document}